\documentclass[
reprint,
amsmath,amssymb,
aps,prl
]{revtex4-2}

\usepackage{graphicx}% Include figure files
\usepackage{dcolumn}% Align table columns on decimal point
\usepackage{bm}% bold math
\usepackage{subfigure}
\usepackage[capitalise]{cleveref}
\UseRawInputEncoding

\begin{document}
	\preprint{APS/123-QED}
	
	\title{Generation of cold polyatomic cations by cascade reactive two-body ion-atom collisions}% Force line breaks with \\
	
	\author{Wei-Chen~Liang$^{1,a}$, Feng-Dong~Jia$^{1,a}$, Fei~Wang$^{2,a}$}
	\thanks{$^a$These authors have contributed equally to this work.}
	
	\author{Yu-Han~Wang$^1$}
	
	\author{Xi~Zhang$^1$}
	
	\author{Jing-Yu~Qian$^1$}
	
	\author{Xiao-Qing~Hu$^3$}
	
	\author{Yong~Wu$^3$}
	
	\author{Jian-Guo~Wang$^3$}
	\email{wang\_jianguo@iapcm.ac.cn}
	
	\author{Ping~Xue$^2$}
	\email{xuep@tsinghua.edu.cn}
	
	\author{Zhi-Ping~Zhong$^1$}
	\email{zpzhong@ucas.ac.cn}
	
	\affiliation{$^1$School of Physical Sciences, University of Chinese Academy of Sciences, Beijing 100049, China}
	\affiliation{$^2$State Key Laboratory of Low-Dimensional Quantum Physics, Department of Physics, Tsinghua University, Beijing 100084, China}
	\affiliation{$^3$Institute of Applied Physics and Computational Mathematics, Beijing 100088, China}
	\date{\today}% It is always \today, today,
	% but any date may be explicitly specified
	
	\begin{abstract}
		
		Polyatomic cations $^{87}$Rb$_M^+$ ($M$ = 2, 3,$\ldots$) have been produced by cascade two-body ion-atom reactive collisions in the two-step CW-laser photoionization of laser-cooled $^{87}$Rb atoms and accumulated in the ion trap. Using resonant-excitation mass spectrometry and resonant excitation-assisted time-of-flight mass spectrometry, we directly observed and distinguished the charged reaction products. We experimentally verified the cascade generation and cascade dissociation of $^{87}$Rb$_M^+$. The populations of $^{87}$Rb$_M^+$ are quantitatively investigated by solving the rate equations. The $^{87}$Rb$^+$-$^{87}$Rb reaction rate coefficient was derived as 9.10$\times10^{-11}$ cm$^3$/s accordingly. The methods developed here for assembling and detecting homonuclear polyatomic cations can be applied to any experiment in ion-atom hybrid traps. The present study lays the foundation for exploring atomically precise metal clusters and physics from few- to many-body perspective.

		\begin{description}
			\item[PACS numbers]
			34.50.Lf,34.90.+q,37.10.Ty,36.40.Wa,36.40.-c
			%May be entered using the \ve$^{87}$Rb$^+$\pacs{#1}+ command.
			%use the optional argument of the \ve$^{87}$Rb$^+$\item+ command to give the category of each item.
		\end{description}
	\end{abstract}
	
	\pacs{Valid PACS appear here}% PACS, the Physics and Astronomy
	\maketitle
	%\tableofcontents
	
	%\section{Introduction}
	
	\textit{Introduction} - Charged-neutral cold chemistry has emerged as a frontier in physical and chemical research, thanks to advances in laser cooling and charged particle trapping technologies \cite{puri_reaction_2019,heazlewood_2021}. The long-range ion-atom interaction is characterized by a $\propto r^{-4}$ polarization potential, where $r$ is the internuclear distance \cite{rmp2019}. This potential results in a large ion-atom interaction cross-section \cite{krukow_reactive_2016}. Molecular ions, as products of ion-atom reactions, are advantageous for trapping and detection due to their additional rotational and vibrational degrees of freedom compared to atomic ions \cite{tomza-2015,daSilvaJr_2015,smialkowski_interactions_2020}. With increasing atom numbers, polyatomic cations can form ion clusters, which bridge individual particles and bulk materials, facilitating the creation of novel materials and devices \cite{heer-clusterrmp-1993,guo-superatom-2022}. However, no polyatomic cations produced in the lab have yet resulted from ion-atom collisions \cite{jacox-polyatomic-2003,boustani-cluster-book,brechignac_charge_2000}. The coldest metal clusters are only generated at 0.37 K using helium nanodroplets\cite{coldcluster-2007}.
	
	Cold ion-atom systems in hybrid traps offer a promising platform for studying interactions and chemical reactions in the millikelvin (mK) regime \cite{heazlewood_2021,puri_reaction_2019,rmp2019}. Previous experiments with these systems have predominantly produced diatomic molecular ions RbCa$^+$ \cite{hall_ion-neutral_2013, hall_light-assisted_2011}, RbBa$^+$ \cite{hall_light-assisted_2013}, CaYb$^+$ \cite{rellergert_measurement_2011}, Ca$_2^+$ \cite{sullivan-Ca2+-2011}, and CaBa$^+$ \cite{sullivan_role_2012} via two-body ion-atom collisions. Ion-atom-atom three-body collisions play a role in producing Rb$_2^+$\cite{dieterle_inelastic_2020,dieterle_transport_2021}, RbBa$^+$\cite{krukow_energy_2016, mohammadi_life_2021}, in which a single ion was immersed in a cloud of ultracold atoms. To date, two-body reaction-rate coefficients have only been reported for a few systems: RbBa$^+$ \cite{hall_light-assisted_2013, krukow_reactive_2016, krukow_energy_2016}, RbCa$^+$ \cite{hall_ion-neutral_2013, hall_light-assisted_2011}, and CaBa$^+$ \cite{sullivan_role_2012}.
	
	Considering that the reactive two-body ion-atom collisions are usually explained by the classical Langevin-capture model \cite{langevin, levine-2009}, where the ion can manifest as either atomic or molecular ion. It is expected that polyatomic cations can also be produced in progressive two-body, reactive ion-atom collisions, whereby each subsequent reaction occurs based on the reaction products formed in the previous step if there are a sizable number of reactant ions and atoms. Within the hybrid trap, the ionic and atomic clouds spatially overlap, and the large ion-atom collision cross-section ensures that the co-trapped ions and atoms can collide within an adequately short range for reactions to proceed. The magneto-optical trap (MOT) maintains a sustained supply of cold atoms, while the continuous wave (CW) ionizing laser continuously generates ions, facilitating the cascade production of $^{87}$Rb$_M^+$. Moreover, the ion trap enables the accumulation and discrimination of charged reaction products.
	
	Herein, we report the creation of a series of polyatomic cations $^{87} \text{Rb}_M^+$ ($M$ = 2, 3, $\ldots$) using continuous-wave (CW) laser photoionization of cold $^{87} \text{Rb}$ atoms. We employed a combination of time-of-flight (TOF) mass spectrometry and resonant-excitation mass spectrometry (REMS) to directly observe the production of $^{87} \text{Rb}_M^+$ and to identify these reaction products. By varying the interaction time of ion-atom collisions and the storage time of ions in the ion trap, we observed a cascade from diatomic cations to polyatomic cations, as well as the cascade dissociation of polyatomic cations.
	
	\begin{figure*}[htbp]%
		\centering
		\vspace{-0.5cm}
		\subfigure[]{\includegraphics[width=0.36\textwidth]{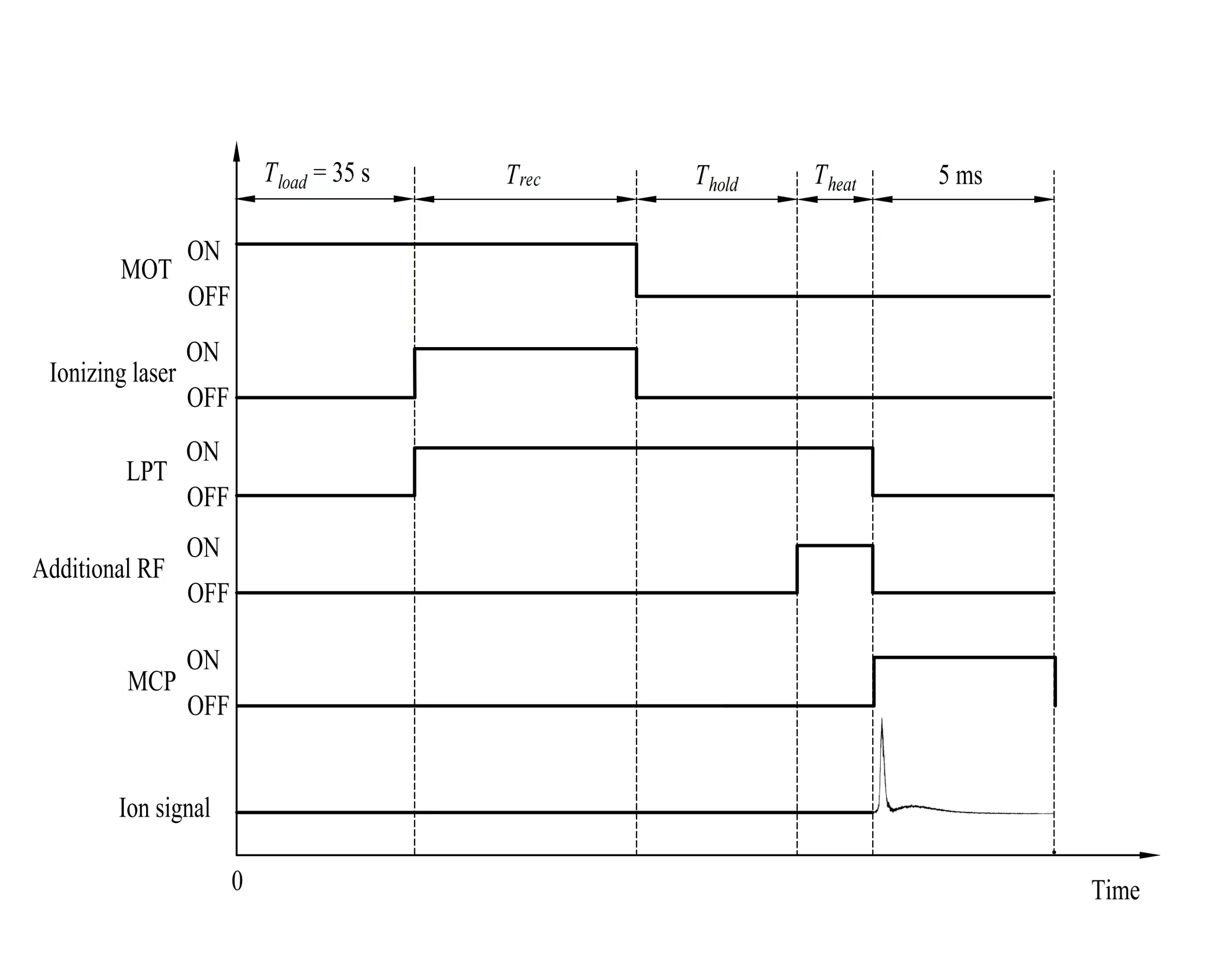}\label{seq}}
		\subfigure[]{\includegraphics[width=0.36\textwidth]{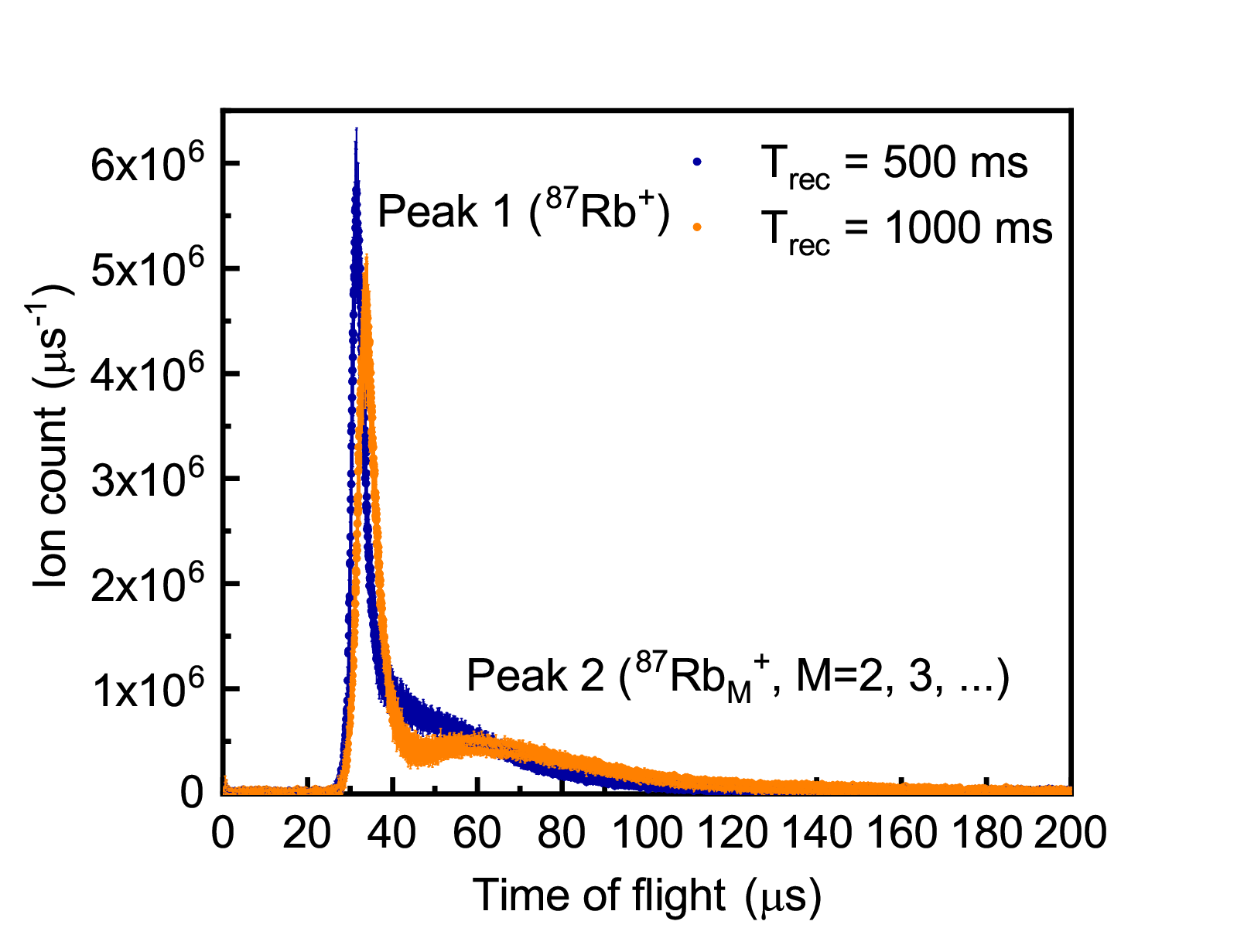}\label{conv-tof}}
		\vspace{-0.3cm}
		\caption{(a) The present experimental sequence diagram for the detection methods. (b) The conventional time of flight (TOF) spectrum (without applying the additional RF field) measured at $T_{\text{rec}}$ = 500 ms (blue) and 1000 ms (orange) with $T_{\text{hold}}=$ 0 ms. $T_{\text{rec}}$ is the duration of ion-atom reactions. The error bars represent the standard error of the mean of ten experiment repetitions. In this and subsequent figures, the ion trap operated at a frequency of $\Omega_{\text{trap}}=2\pi\times550$ kHz, the amplitude of the additional RF field $V_{\text{RF}}=$ 4 V, and the wavelength and intensity of the ionizing laser were 478.8 nm and 530.5 mW/cm$^2$ unless specified. }
		\vspace{-0.4cm}
	\end{figure*} 
	
	\textit{Experimental Setup} - In our experiments, we used an $^{87}$Rb$^+$-$^{87}$Rb hybrid trap, which comprises a standard magneto-optical trap (MOT) and a mass-selective linear Paul trap (LPT)\cite{lv_measurement_2017}. The sequence of our experiments is shown in Fig.\ref{seq}. Specifically, the MOT was first loaded to a steady state. Then the ionizing laser and the LPT were then switched on simultaneously. The MOT, ionizing laser, and LPT worked together for an adjustable period $T_{\text{rec}}$.  At the end of $T_{\text{rec}}$, the ionizing laser was switched off and the ions were held in the LPT for an adjustable period $T_{\text{hold}}$. The MOT can be switched off or kept on during $T_{\text{hold}}$. After $T_{\text{hold}}$, the MOT was switched off and an additional RF field with a frequency $\Omega_{\text{RF}}$ was applied to the quadrupole electrodes. The additional RF field and LPT worked together for an adjustable period $T_{\text{heat}}$. After $T_{\text{heat}}$, the voltage on the end-cap ring electrode closer to the microchannel plate (MCP) was turned off; this pushed the trapped ions toward the MCP, and the resulting TOF spectrum was recorded by an oscilloscope. The relative ion intensity measured by the MCP was converted into an absolute ion number by combining atomic absorption imaging and rate equations\cite{liang-fit-2022}. The intensity, position, and half-width for a peak in a TOF spectrum were derived by fitting using the Gumbel probability density function \cite{liang-fit-2022}. 
	
	Two approaches were used to discriminate ion species, one is conventional TOF measurements, where the additional RF field was not applied and $T_{\text{heat}}$ was zero in this work. Another is REMS, in which an additional RF field with a frequency $\Omega_{\text{RF}}$ was applied. When $\Omega_{\text{RF}}$ resonates with the radial secular frequency $\omega_r$ (determined by the Mathieu equations\cite{rmp-paultrap,goodman-iontrap-2012}) of an ion species in the LPT, this species of ions are heated and leave the LPT. This causes a reduction in the total number of ions, thus yielding a resonance in the REMS. In our experiments, we measure the REMS by obtaining the total ion count as a function of $\Omega_{\text{RF}}$. 
	
	\begin{figure}[htbp]%
		\centering
		\vspace{-0.2cm}
		\includegraphics[width=0.40\textwidth]{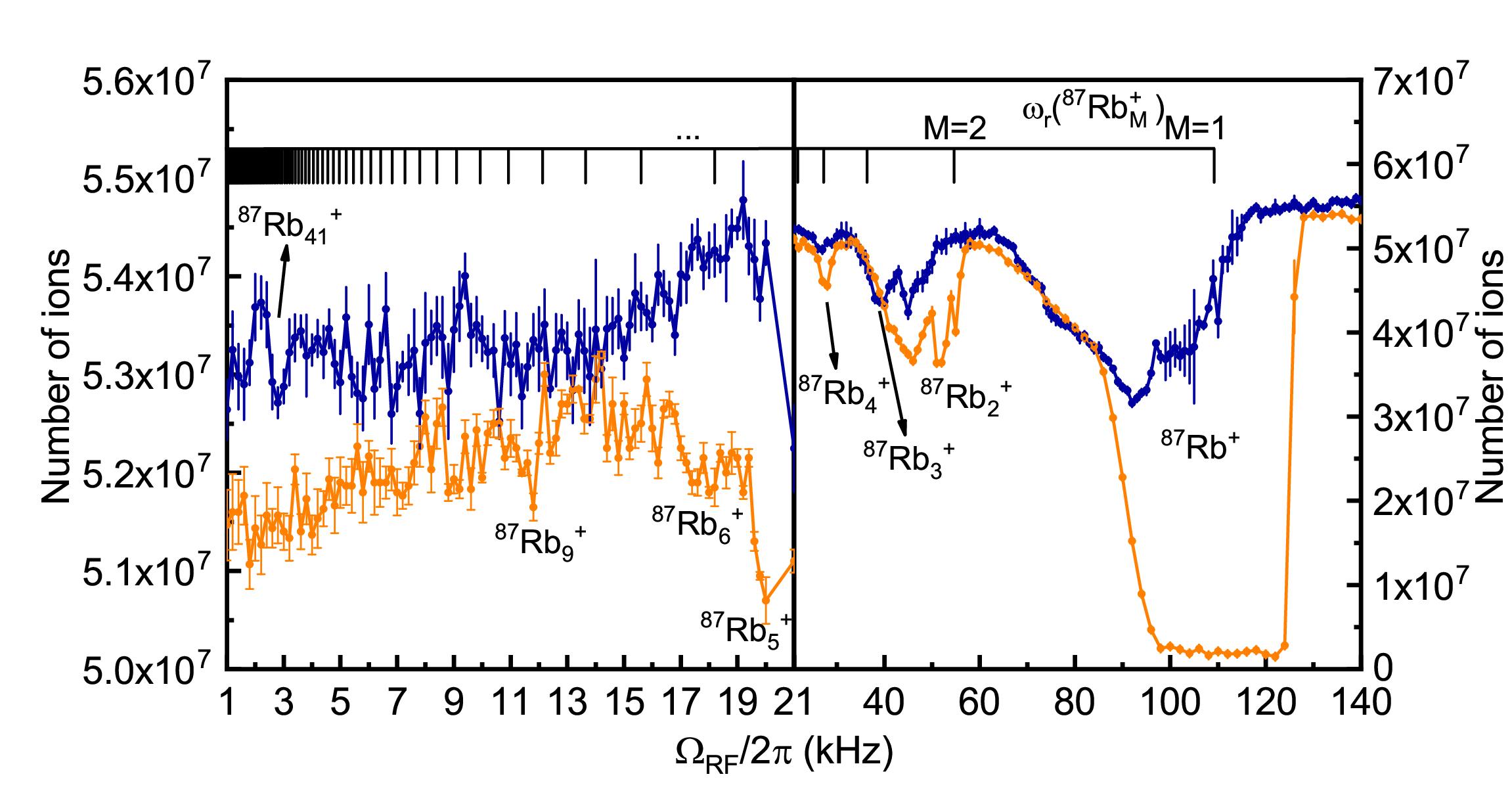}
		\vspace{-0.3cm}
		\caption{The resonant-excitation mass spectra measured at  $T_{\text{hold}}=$ 0 ms (blue) and 800 ms (orange) at $T_{\text{rec}}=$ 750 ms. For the dataset of $T_{\text{hold}}=$ 0 ms, the additional RF field was applied for $T_{\text{heat}}=$ 1 ms in the range of $\Omega_{\text{RF}}=2\pi\times$ 0 to 30 kHz, 250 $\mu s$ in $\Omega_{\text{RF}}=2\pi\times$ 30 to 60 kHz and 150 $\mu s$ in $\Omega_{\text{RF}}=2\pi\times$ 60 to 140 kHz. For the dataset of $T_{\text{hold}}=$ 800 ms, the additional RF field was applied for $T_{\text{heat}}=$ 1 ms. The data in the range of $\Omega_{\text{RF}}=2\pi\times$ 0 to 10 kHz was repeated six times and the data in $\Omega_{\text{RF}}=2\pi\times$ 10$ \sim $150 kHz was repeated four times. The error bars represent the standard error of the mean of multiple experiment repetitions. The theoretical radial secular frequency $\omega_r$ for $^{87}$Rb$_M^+$ marked in the figures (line marker) is calculated as $\omega_r$($^{87}$Rb$_M^+$) = $2\pi\times109.2/M$ kHz according to the Mathieu equations\cite{rmp-paultrap,goodman-iontrap-2012}.}
		\label{rems}
		\vspace{-0.4cm}
	\end{figure}
	
	%\section*{Production of polyatomic cations by sequential ion-atom collisions}\label{sec-create}
	
	\textit{Results} – In the measured conventional TOF spectrum shown in Fig.\ref{conv-tof}, there is a strong $^{87}$Rb$^+$ peak (peak 1) and a broad peak (peak 2). The peak 2 should be products of ion-atom reactions, i.e., $^{87}$Rb$_M^+$  ($M\geq$2). Our simulations suggest that the broadness of peak 2 may result from the overlap of multiple peaks (see Fig. S2 in the Appendices). Additionally, the position of peak 2 increased with $T_{\text{rec}}$, indicating that the effective mass of the ions in the peak 2 has increased.
	
	To distinguish the coexisting ion species in the ion trap, REMS spectra were measured at $T_{\text{hold}} = $0 ms and 800 ms after $T_{\text{rec}} = 750$ ms. In these measurements, an additional RF field was applied for different $T_{\text{heat}}$ over a range of $\Omega_{\text{RF}}$ to ensure that the resonances correspond to signals of $^{87}\text{Rb}_M^+$ rather than fractional resonances (see Fig. S4 in the Appendices). As shown in Fig. \ref{rems}, the positions of most resonances were generally in good agreement with the theoretical predictions for $\omega_r$($^{87}\text{Rb}_M^+$) (marked in Fig.\ref{rems}). Some resonances slightly deviated from their theoretical positions due to the perturbation of the radial secular motion by interactions between the trapped ions in the LPT \cite{goodman-iontrap-2012, blumel-iontrap-1989, chu-iontrap-1998}. Besides, many resonances were located at the second harmonic frequencies $2\omega_r$ of specific ion species. This is likely due to side-effect heating of the ions by the additional RF field\cite{sivarajah-iontrap-2012, goodman-iontrap-2012, razvi-fractional-1998}. Notably, at $T_{\text{hold}} = 0$ ms, an observable signal of $^{87}\text{Rb}_{41}^+$ was detected.
	
	To quantitatively interpret our experimental results, the rate equations of ions were established following the classical Langevin-capture model \cite{langevin, levine-2009}. We assume the observed polyatomic ions can only be produced through two-body ion-atom radiative associations (RA) due to low MOT atomic density\cite{rmp2019}. 
	
	\begin{figure}[htbp]%
		\centering
		\vspace{-0.2cm}
		\subfigure{\includegraphics[width=0.23\textwidth]{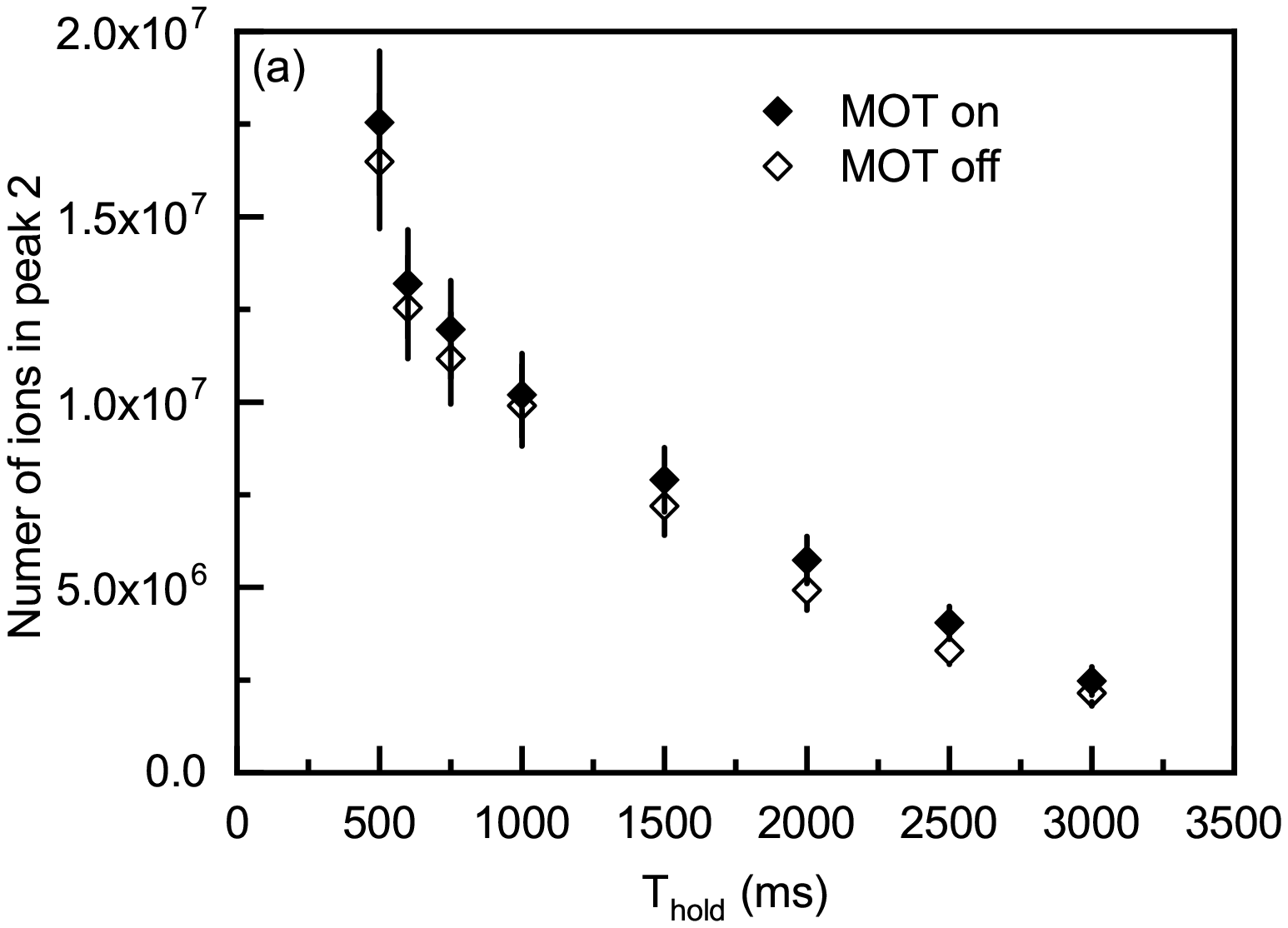}\label{lptmot}}
		\subfigure{\includegraphics[width=0.23\textwidth]{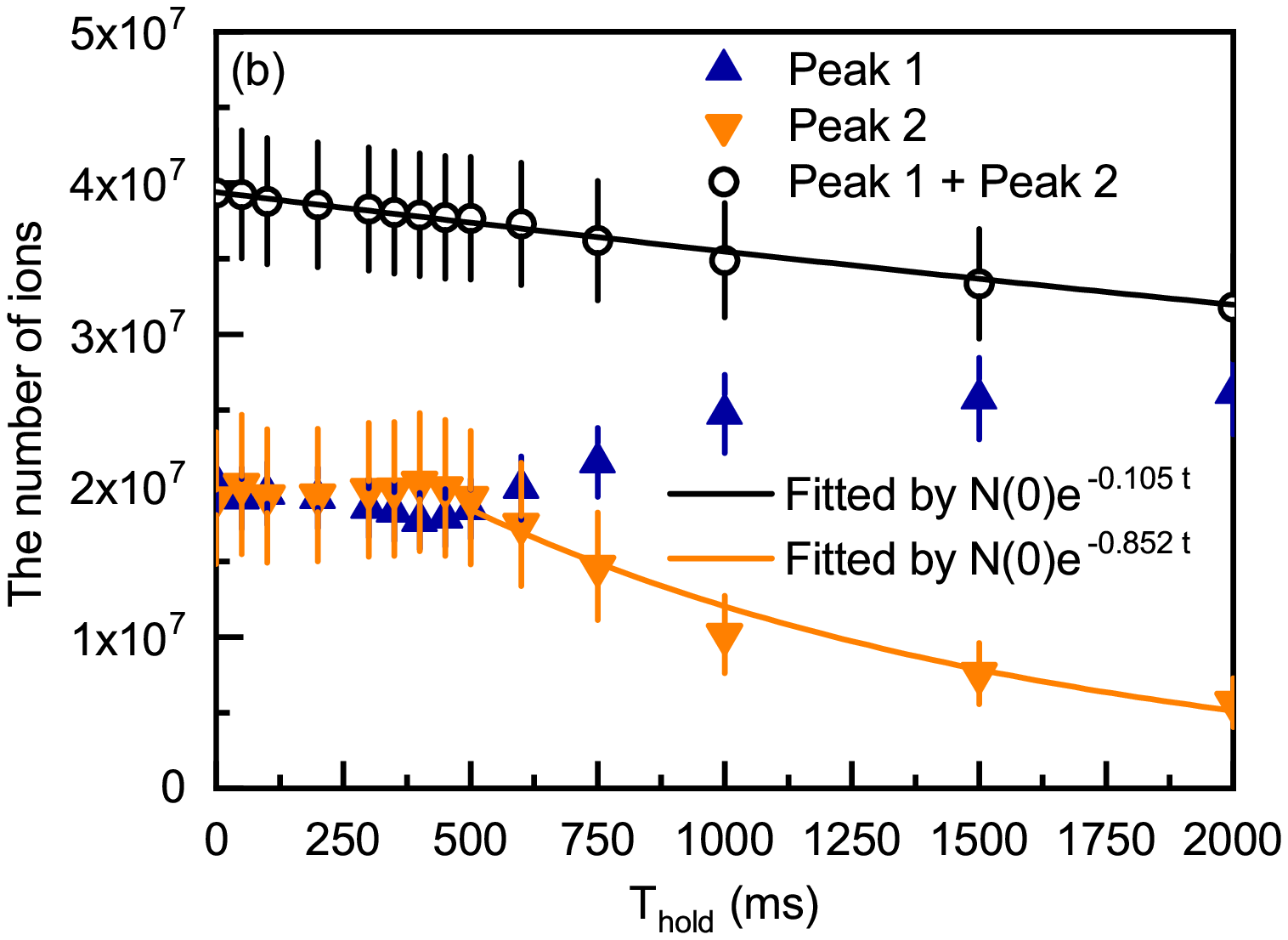}\label{lpt}}
		\vspace{-0.3cm}
		\caption{(a) The measured number of ions in the peak 2 as a function of $T_{\text{hold}}$ with (black solid diamond) or without (black hollow diamond) the presence of cold atoms after $T_{\text{rec}}=$ 750 ms. The error bars represent the standard error of the mean of three experiment repetitions. (b) Number of ions in the peaks 1 (blue triangles) and 2 (orange reversed triangles) and the total number of ions (black hollow circles) measured as a function of $T_{\text{hold}}$ after $T_{\text{rec}}=$ 750 ms. The ionizing laser was kept off during $T_{\text{hold}}$. The error bars represent the standard error of the mean of three experiment repetitions.}
		\vspace{-0.3cm}
	\end{figure}
	
	Since $^{87}\text{Rb}^+$ ions are generated by photoionizing $^{87}\text{Rb}$ atoms in the MOT, the initially produced ion cloud inherits the spatial dimensions of the MOT atomic cloud. Subsequently, the ion cloud rapidly expands to a larger bulk due to the significantly greater volume of the LPT compared to that of the MOT \cite{li_rb_2020}. In our experiments, the ions trapped in the LPT exhibit minimal interaction with atoms, as illustrated in Fig.\ref{lptmot} because the ionic density in the LPT is substantially lower than in the MOT \cite{li_rb_2020}. Consequently, ion-atom reactions can be neglected when the ionizing laser is turned off, and the duration of the ionizing laser, $T_{\text{rec}}$, corresponds to the ion-atom reaction time. Therefore, the rate equation for the ions can be separated into two components: the rate equation for ions in the MOT and the rate equation for ions in the LPT.
	
	There are two main mechanisms leading to ion loss in the linear Paul trap (LPT): limited trapping capability and dissociation of polyatomic ions that may be caused by the oscillating electric field of the Paul trap\cite{perez-rios_electric-field_2021}. As shown in Fig.\ref{lpt}, the total number of ions displays a single exponential decay, indicating uniform trapping for all ion species. The ion loss rate $\Gamma_t$ in the LPT was found to be 0.11 s$^{-1}$. However, for $T_{hold} \geq 500$ ms, ions in peak 2 decay faster than the total number of ions, suggesting that polyatomic ions undergo additional losses due to dissociation. Assuming uniform dissociation rates among all polyatomic ions, their dissociation rate was estimated to be 0.75 s$^{-1}$, based on the decay rate of ions in peak 2 at $T_{hold} \geq 500$ ms. Meanwhile, the intensity of peak 1, representing $^{87}$Rb$^+$, increases with $T_{hold}$ due to the interplay between polyatomic ion dissociation and the finite lifetime of $^{87}$Rb$^+$ in the LPT. For $T_{hold} \geq 500$ ms, the peak 1's intensity rises slowly while the peak 2 decays. This pattern suggests that the dissociation of $^{87}$Rb$_2^+$ contributes to the gradual decline of peak 2's intensity, as it predominantly results in the production of $^{87}$Rb$^+$. This implies that polyatomic ions dissociate in a cascade manner.
	
	According to these analysis, the rate equations for $^{87}$Rb$^+$ and $^{87}$Rb$_M^+$ ($M\geq2$) are given as follows.
	
	\begin{equation}\label{ionmot}
		\begin{aligned}
			&\frac{dN_{Rb^+,MOT}}{dt}=\gamma _{PI}N_{Rb,5P}+\Gamma _{diss,Rb_2^+}N_{Rb_2^+,MOT}\\
			&-k_2n_{Rb,5P}N_{Rb^+,MOT}-\left( \Gamma _{diff,Rb^+}+\Gamma _t \right) N_{Rb^+,MOT},
		\end{aligned}
	\end{equation}
	
	\begin{equation}\label{ionlpt}
		\begin{aligned}
			&\frac{dN_{Rb^+,LPT}}{dt}=\Gamma _{diff,Rb^+}N_{Rb^+,MOT}+\Gamma _{diss,Rb_2^+}N_{Rb_2^+,LPT}\\
			&-\Gamma _tN_{Rb^+,LPT},
		\end{aligned}
	\end{equation}
	
	\begin{equation}\label{molmot}
		\begin{aligned}
			&\frac{dN_{Rb_{M}^+,MOT}}{dt}=k_{2}(Rb_{M-1}^+)n_{Rb,5P}N_{Rb_{M-1}^+,MOT}\\
			&-k_{2}(Rb_M^+)n_{Rb,5P}N_{Rb_M^+,MOT}+\Gamma _{diss,Rb_{M+1}^+}N_{Rb_{M+1}^+,MOT}\\
			&-\left( \Gamma _{diss,Rb_{M}^+}+\Gamma _{diff,Rb_M^+}+\Gamma _t \right) N_{Rb_M^+,MOT},
		\end{aligned}
	\end{equation}
	
	\begin{equation}\label{mollpt}
		\begin{aligned}
			&\frac{dN_{Rb_M^+,LPT}}{dt}=\Gamma _{diff,Rb_M^+}N_{Rb_M^+,MOT}\\
			&+\Gamma _{diss,Rb_{M+1}^+}N_{Rb_{M+1}^+,LPT}-\left( \Gamma _t+\Gamma _{diss,Rb_M^+} \right) N_{Rb_M^+,LPT}
		\end{aligned}
	\end{equation}
	
	Here, only $^{87}$Rb atoms in the $5P_{3/2}$ state were considered. This is because the RA cross sections for $^{87}$Rb($5P_{3/2}$) are $10^3$ times larger than those for $^{87}$Rb($5S_{1/2}$), due to significantly stronger dipole interactions, as indicated by our full quantum-mechanical calculations\cite{yw-2024}. The number of atoms in the $5P_{3/2}$ state, $N_{Rb,5P}$, is given by $N_{Rb,5P}(t) = f_{5P} \times (A e^{-\gamma_x t} + N_e)$\cite{liang-fit-2022,wang-atomion-2022}, where $A$, $\gamma_x$, and $N_e$ were determined by absorption imaging\cite{lv_measurement_2017}. The photoionization rate $\gamma_{PI}$ is estimated to be 2.43 s$^{-1}$. The fraction of $^{87}$Rb atoms $f_{5P}$ in the $5P_{3/2}$ state is approximately 0.32\cite{lv_measurement_2017,li_rb_2020}. The ion loss rate due to the limited trapping capability of the LPT, $\Gamma_t$, and the dissociation rate of $^{87}$Rb$_M^+$, $\Gamma_{diss,Rb_{M}^+}$, are estimated to be 0.11 s$^{-1}$ and 0.75 s$^{-1}$, respectively, as discussed earlier.
	
	For the diffusion rates, we assume that the diffusion rates for all ions are identical. The diffusion rate is proportional to the kinetic energy (temperature) of the ions in the MOT. Based on energy and momentum conservation principles, the initial temperature of $^{87}$Rb$^+$ ions generated by the photoionization of cold atoms is estimated to be a few millikelvin (mK). Unlike ultracold neutral plasmas formed by photoionizing cold atoms, where electron-ion interactions typically raise the ion temperature to a few Kelvin\cite{killian-unp-2007}, the ions in this work are expected to remain much cooler due to the repulsion of electrons by the LPT. Consequently, we choose a diffusion rate coefficient of 0.1 s$^{-1}$, corresponding to an ion temperature of approximately 10 mK. Notably, varying the diffusion rate between 0.001 s$^{-1}$ and 1.0 s$^{-1}$ has negligible impact on the calculation results.
	
	$k_{2}$($^{87}$Rb$_M^+$) ($M\geq1$) are reaction-rate coefficients for $^{87}$Rb$_M^+$-$^{87}$Rb. Based on the Langevin model\cite{langevin, levine-2009}, $k_{2}$($^{87}$Rb$_M^+$) is proportional to $\sqrt{C_4/\mu_M}$. Here, $C_4$ is proportional to the static electric dipole polarizability of $^{87}$Rb, and $\mu_M$ is the reduced mass of the colliding $^{87}$Rb$_M^+$ and $^{87}$Rb. By fitting the measured evolution of the ion numbers in the peaks 1 and 2 as a function of $T_{\text{rec}}$ using the rate equations (shown in Fig.\ref{cal-convtof}), we derived $k_{2}$($^{87}$Rb$^+$) to be 9.10$\times10^{-11}$ cm$^3$/s (see Appendices for discussion). This value is comparable to the two-body ion-atom reaction-rate coefficient of $7\times10^{-11}$ cm$^3$/s reported in the production of RbBa$^+$\cite{hall_light-assisted_2013}. Therefore, $k_{2}$($^{87}$Rb$_M^+$) can be approximated by $k_{2}$($^{87}$Rb$_M^+$) = $k_{2}$($^{87}$Rb$^+$)$\times\sqrt{2/\mu_M}$.
	
	\begin{figure*}[htbp]%
		\centering
		\vspace{-0.4cm}
		\subfigure[]{\includegraphics[width=0.28\textwidth]{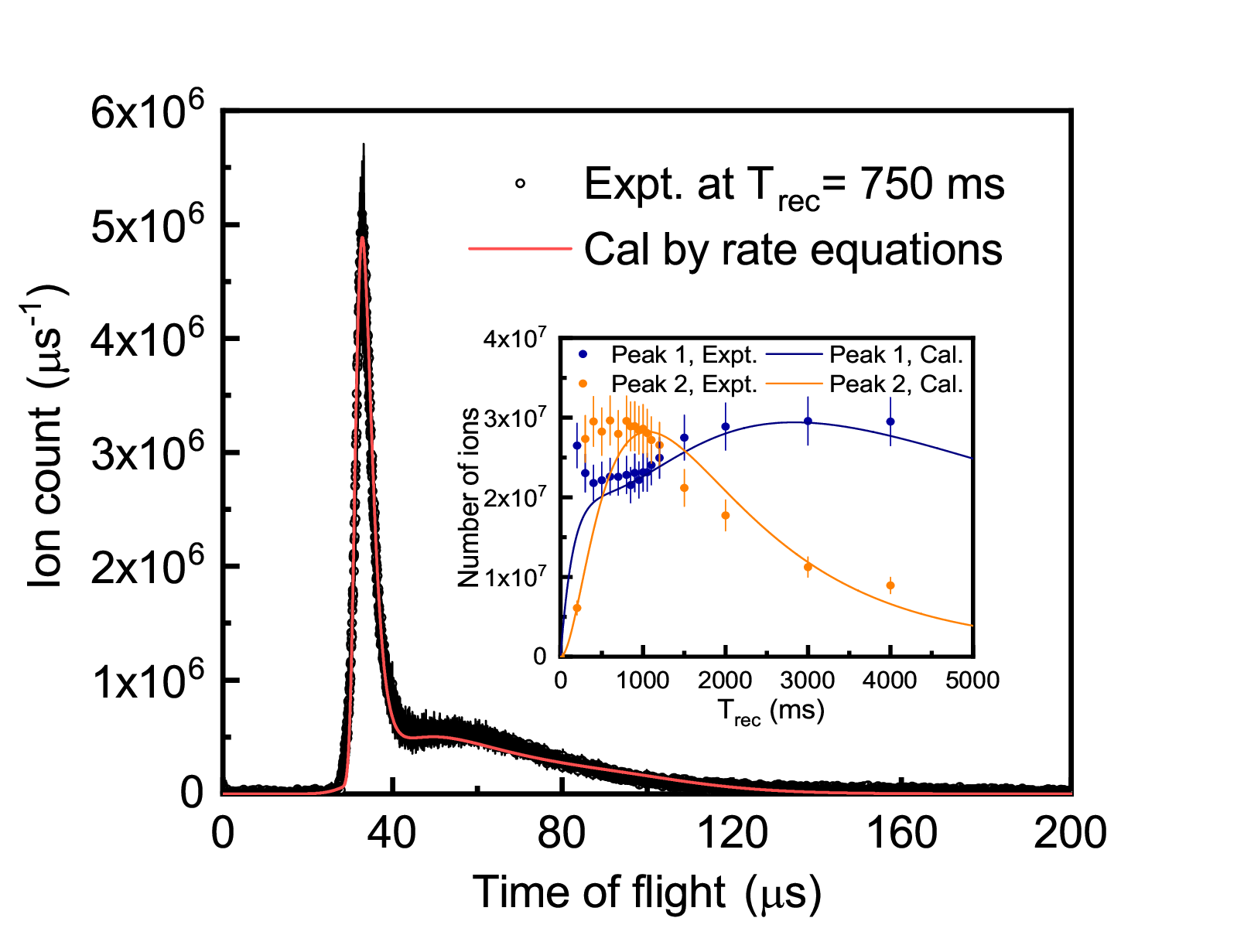}\label{cal-convtof}}
		\subfigure[]{\includegraphics[width=0.28\textwidth]{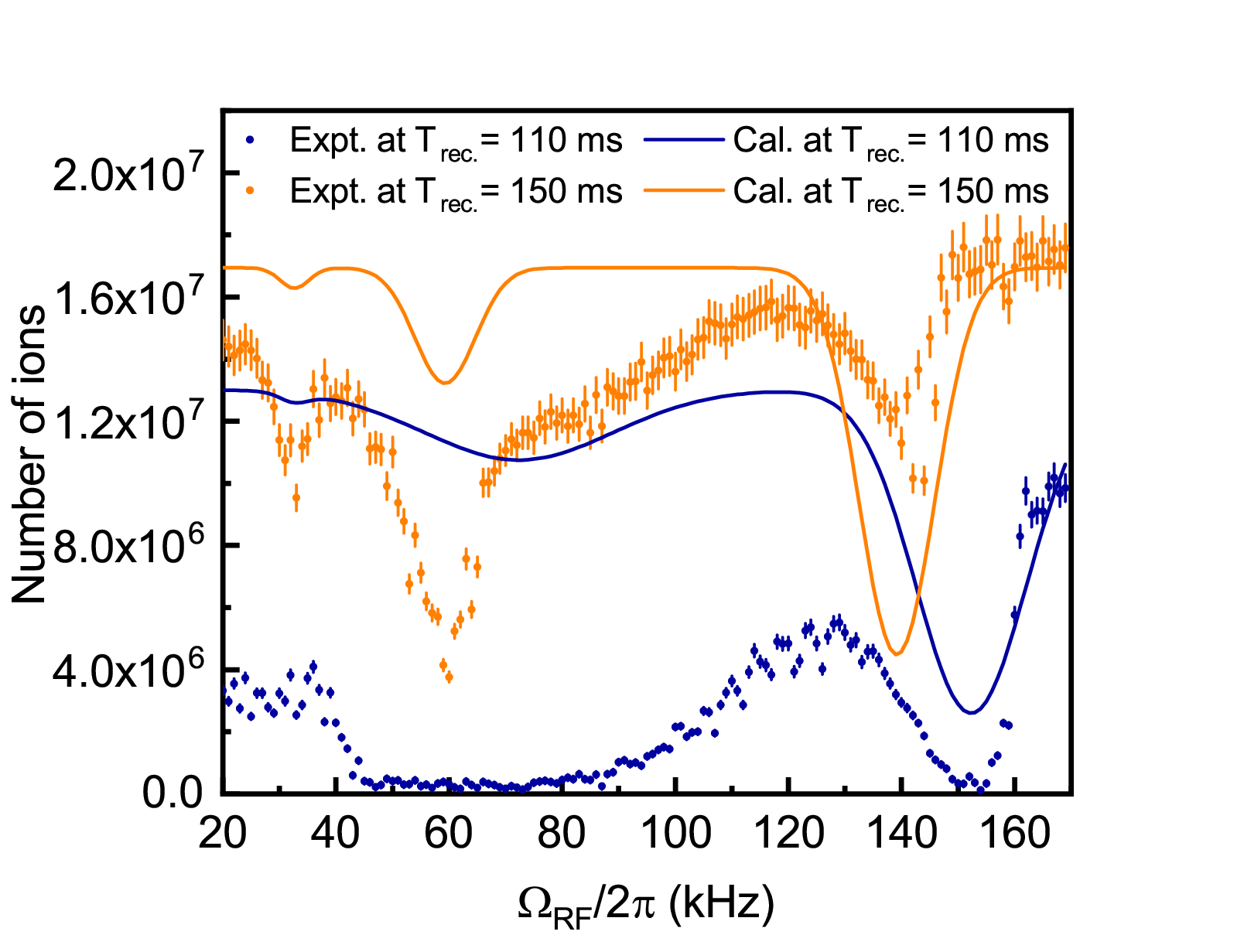}\label{rems-prod}}
		\subfigure[]{\includegraphics[width=0.28\textwidth]{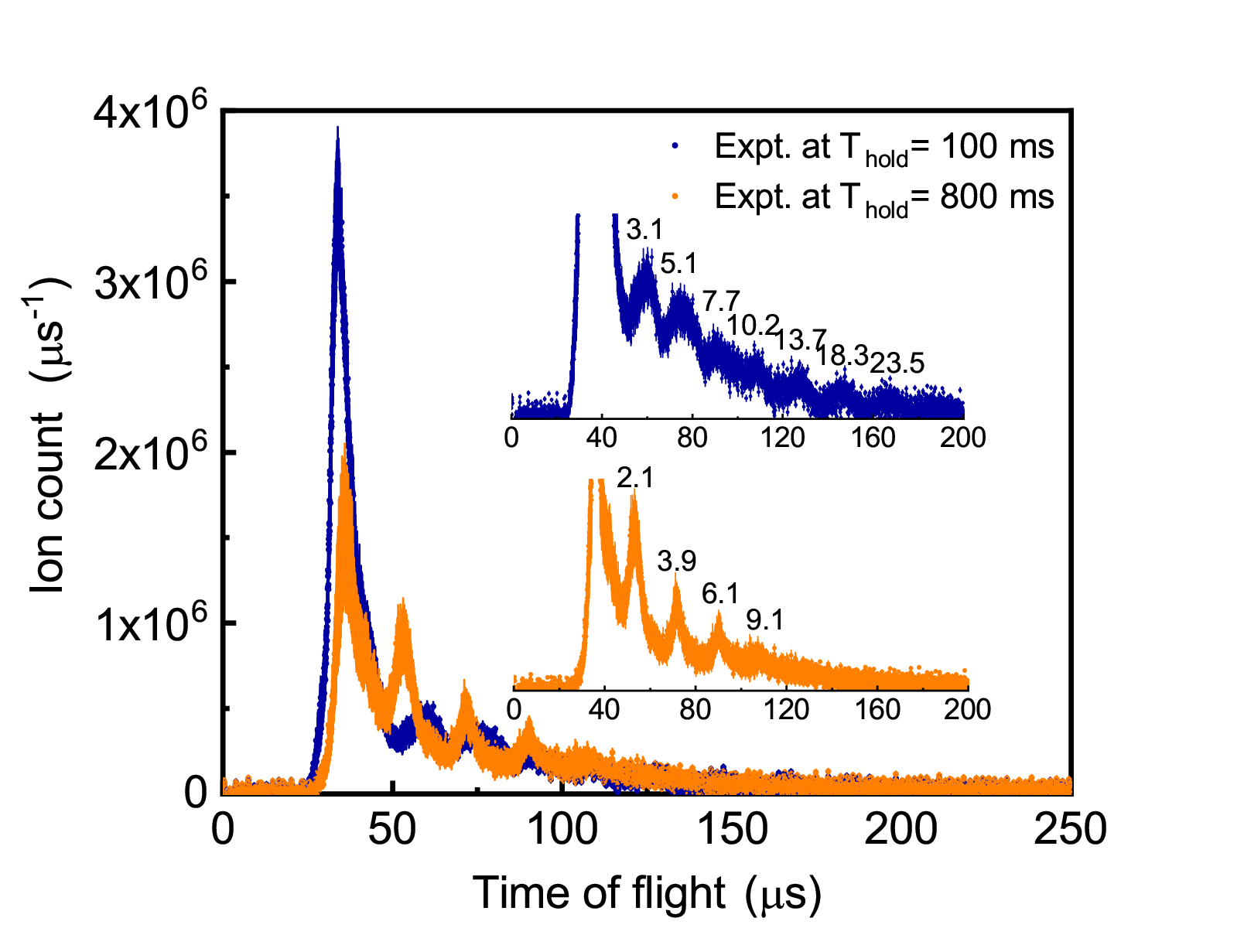}\label{re-tof}}
		\vspace{-0.4cm}
		\caption{(a) Comparison of the measured conventional TOF spectrum at $T_{\text{rec}} = 750$ ms (black hollow dots) and calculated TOF spectrum using rate equations (red line). The inset compares the measured (dots) and simulated (solid lines) evolution of ion numbers in the peak 1 (dark blue) and peak 2 (orange) with $T_{\text{rec}}$ using the rate equations. Error bars represent the standard error of the mean from three experimental repetitions. In the calculated TOF spectrum, the line shape of each peak was described by the Gumbel distribution \cite{liang-fit-2022}. The amplitude of the $^{87}\text{Rb}^+$ and $^{87}\text{Rb}_M^+$ peaks were determined by the calculated number of ions. The width of each peak was derived from our simulated TOF spectrum (see Fig. S2 in the Appendices). The position of each $^{87}\text{Rb}_M^+$ peak was calculated using the squared mass ratio $\sqrt{M}$ relative to $^{87}\text{Rb}^+$, with the $^{87}\text{Rb}^+$ position obtained from the peak 1 in the measured TOF spectrum. (b) Comparison of the measured and calculated resonant-excitation mass spectrum as a function of $T_{\text{rec}}$. The blue and orange dots represent signals measured at $T_{\text{rec}} = 110$ ms and $T_{\text{rec}} = 150$ ms, respectively. And the blue and orange solid lines represent the calculated spectra at $T_{\text{rec}} = 110$ ms and $T_{\text{rec}} = 150$ ms, respectively. During the experimental measurements, an additional RF field was applied for $T_{\text{heat}} = 170$ $\mu$s. The error bars represent the standard error of the mean from four experimental repetitions. In each calculated spectrum, the depth of a resonance was determined by the number of $^{87}\text{Rb}_M^+$ ions calculated using the rate equation as described in the main text. The position, width, and shape of each resonance were determined by the corresponding experimental data. (c) RE-TOF spectrum measured at $T_{\text{hold}} = 100$ ms (blue) and $T_{\text{hold}} = 800$ ms (orange) after $T_{\text{rec}} = 750$ ms. The ionizing laser and MOT were kept off during $T_{\text{hold}}$. In each spectrum, an additional RF field with $\Omega_{\text{RF}} = 2\pi \times 109.2$ kHz was applied for $T_{\text{heat}} = 170$ $\mu$s. The marked numbers above each split peak in the insets represent the effective number of atoms contained in the corresponding polyatomic ion, which was calculated by [$T/T$(peak 1)]$^2$, where $T$ and $T$(peak 1) are the positions of the corresponding peak and peak 1, respectively. The error bars represent the standard error of the mean from twenty experimental repetitions.}
		\vspace{-0.4cm}
	\end{figure*}
	
	With all the parameters of the rate equations determined, the number of ions at specific $T_{\text{rec}}$ can be calculated. Accordingly, we simulated the TOF spectrum at $T_{\text{rec}}=$ 750 ms based on the calculations. The calculated and measured TOF spectrum are in resonably good agreement, as shown in Fig.\ref{cal-convtof}.
	
	Our calculation suggested that the number of $^{87}$Rb$_M^+$ with $M\geq10$ should be too small to identify at $T_{\text{rec}}=$ 750 ms. Beside the observable signal for $^{87}$Rb$_{41}^+$ in REMS measured at $T_{\text{hold}}=$ 0 ms, as shown in Fig.\ref{rems}, other observed $^{87}$Rb$_{M}^+$ are generally $M\leq9$. The processes in the generation of these large polyatomic cations are to be further investigated.
	
	To validate the cascade production of these reaction products $^{87}$Rb$_M^+$, we measured the REMS as a function of $T_{\text{rec}}$. Fig.\ref{rems-prod} displays REMS at reaction times of 110 ms and 150 ms, respectively. Two resonances (for $^{87}$Rb$^+$ and $^{87}$Rb$_2^+$) were observed at $T_{\text{rec}}=$ 110 ms. As $T_{\text{rec}}$ increased to 150 ms, three resonances were generated following the addition of $^{87}$Rb$_3^+$. As $T_{\text{rec}}$ increased to 750 ms, resonances were detected at lower excitation frequencies, as shown in Fig.\ref{rems}, indicating the production of polyatomic ions containing more atoms. We have simulated the corresponding REMS according to the calculations by rate equations. As shown in Fig.\ref{rems-prod}, our calculations can reveal the cascade generation of polyatomic ions, which is consistent with experimental characteristics. However, there is a significant difference between the calculated ion strength distribution and the experiment, especially for $T_{\text{rec}}=$ 110 ms. The reason to this may be attributed to that different types of ions in LPT have different responses to the excitation RF field, similar to spectral response of detectors. In our experiments, the additional RF field was applied for a fixed duration and amplitude in the measurement of REMS. As a result, the depth of a resonance can hardly reflect the accurate number of corresponding ion species. 
	
	To investigate the dissociation characteristics of polyatomic ions, we employed resonant excitation-assisted time-of-flight (RE-TOF) spectrometry, in which an additional RF field was applied to discriminate polyatomic ions in the peak 2 (see Appendices for details). We measured RE-TOF spectra as a function of $T_{hold}$ at $T_{\text{rec}} = 750$ ms. As shown in Fig.\ref{re-tof}, at $T_{hold} = 100$ ms, seven discrete peaks were observed alongside the $^{87}$Rb$^+$ peak. At $T_{hold} = 800$ ms, four discrete peaks were observed alongside the $^{87}$Rb$^+$ peak. Each split peak was assigned using the squared ratio of its position $T$ to the position of the $^{87}$Rb$^+$ peak $T$(peak 1), i.e., [$T/T$(peak 1)]$^2$, which represents the effective number of atoms contained in the corresponding polyatomic ion. Note that the values of [$T/T$(peak 1)]$^2$ deviate from integers, as each split peak likely contains contributions from multiple ion species. This arises due to the limited resolution of the mass spectrometer and the odd-even staggering observed in the intensity distribution of $^{87}$Rb$_M^+$ with $M$\cite{boustani-cluster-book}. As $T_{hold}$ increased from 100 ms to 800 ms, the highest observed $M$ for $^{87}$Rb$_M^+$ decreased from approximately 25 to 9. 
	
	In the REMS measurements shown in Fig.\ref{rems}, a signal was observed at the position of $^{87}$Rb$_4^+$ for $T_{hold} = 0$ ms, with a stronger signal at $T_{hold} = 800$ ms. These observations in both the RE-TOF spectrum and REMS strongly suggest that $^{87}$Rb$_M^+$ ions dissociate in a cascade manner. Additionally, in the RE-TOF spectrum measured at $T_{hold} = 800$ ms, $^{87}$Rb$_M^+$ ions with $M \approx$ 2, 4, 6, and 9 are distinguishable, and these ions also show significant signals in the REMS at $T_{hold} = 800$ ms, as shown in Fig.\ref{rems}. This indicates that RE-TOF spectrometry is a reliable method for analyzing these polyatomic ions.
	
	\textit{Conclusion} - In summary, an abundance of cold polyatomic cations, denoted as $^{87}$Rb$_M^+$ ($M$ = 2, 3,$\ldots$) was generated within an $^{87}$Rb$^+$-$^{87}$Rb mixture created through CW-laser photoionization of cold $^{87}$Rb atoms. The utilization of a magneto-optical trap (MOT) and a CW ionizing laser allowed for the sustained production of cold atoms and ions, while the ion trap facilitated the accumulation and discrimination of charged reaction products. The temperature of the produced polyatomic cations was found to be no greater than 10 mK. These atomically precise polyatomic cations are significant for studying physics from few- to many-body perspectives. By employing the Langevin-capture model\cite{langevin,levine-2009}, we established rate equations and determined the reaction rates for $^{87}$Rb$_M^+$-$^{87}$Rb interactions. These low-temperature reaction rates are instrumental for validating quantum chemical theories. 
	
	Using conventional time-of-flight (TOF) spectrometry and resonant excitation-assisted TOF (RE-TOF) spectrometry, we detected and distinguished the formation of $^{87}$Rb$_M^+$ ions. Our results indicate the occurrence of cascade reactions, leading to a series of polyatomic cations. RE-TOF spectrometry further revealed distinct cascade features in the dissociation of $^{87}$Rb$_M^+$. Our findings provide a comprehensive understanding of the dynamics of cold ion-atom interactions and chemical reactions and open an avenue for controlled cold ion-neutral chemistry, under conditions similar to those found in the interstellar medium.
	
	\begin{acknowledgments}
		
		The authors express their deep appreciation of Dr. Xin-Yu Luo (Max-Planck-Institute for Quantum Optics), Prof. Zhen-Sheng Yuan (University of Science and Technology of China), and Prof. Thomas Gallagher (University of Virginia) for their fruitful discussions.
		
		This study was supported by the National Key Research and Development Program of China (Grant Nos. 2017YFA0402300 and 2017YFA0304900), Beijing Natural Science Foundation (Grant No. 1212014), Fundamental Research Funds for the Central Universities, the Strategic Priority Research Program (Grant No. XDB28000000) of the Chinese Academy of Sciences, specialized research fund for CAS Key Laboratory of Geospace Environment (Grant No. GE2020-01), and the National Natural Science Foundation of China (Grant Nos. 61975091, 61575108).
	\end{acknowledgments}
	
	\bibliographystyle{apsrev}
	\bibliography{polyatomic}
	
\end{document}